\documentclass{osa-article}

\usepackage{graphicx}
\usepackage{float}
\usepackage{dcolumn}
\usepackage{relsize}
\usepackage{amsmath}
\usepackage{bm}
\usepackage{xcolor}
\usepackage[hang]{subfigure}
\usepackage{dsfont}

\journal{osajournal}
\articletype{Research Article}

\begin{document}

\title{Husimi Functions for Coupled Optical Resonators}

\author{Martí Bosch,\authormark{1} Arne Behrens,\authormark{2}, Stefan Sinzinger \authormark{2}, and Martina Hentschel\authormark{3}}

\address{\authormark{1}Institut f{\"u}r Physik, Technische Universität Ilmenau, D-98693 Ilmenau, Germany\\
\authormark{2}Fachgebiet Technische Optik, Institut für Mikro- und Nanotechnologien MacroNano, Technische Universität Ilmenau, D-98693 Ilmenau, Germany\\
\authormark{3}Institute of Physics, Technische Universität Chemnitz, D-09107 Chemnitz, Germany}

\email{\authormark{*}marti.bosch@tu-ilmenau.de} %% email address is required

\begin{abstract}
Phase-space analysis has been widely used in the past for the study of optical resonant systems. While it is usually employed to analyze the far-field behaviour of resonant systems we focus here on it's applicability to coupling problems. By looking at the phase-space description of both the resonant mode and the exciting source it is possible to understand the coupling mechanisms as well as to gain insights and approximate the coupling behaviour with reduced computational efforts. In this work we develop the framework for this idea and apply it to a system of an asymmetric dielectric resonator coupled to a waveguide.  
\end{abstract}

%%%%%%%%%%%%%%%%%%%%%%%%%%  body  %%%%%%%%%%%%%%%%%%%%%%%%%%
\section{Introduction}
Over the last years, optical microcavities have attracted researcher's attention in both fundamental and applied physics due to their high quality factor (Q) and small mode volumes \cite{Vahala.2003}. Besides applications in sensing \cite{Foreman.2015}, nonlinear optics \cite{Ilchenko.2004}, light-matter interaction \cite{Xiao.2012} and lasing \cite{Harayama.2011}, they have been used as a research platform to study exciting physical phenomena such as exceptional points \cite{Wiersig.2014,Chen.2017} and optical chirality \cite{Liu.2018}. In recent years, the application range of optical microresonators has been further expanded by introducing systems of deformed or perturbed resonators. This leads to new effects which can be used for applications such as microlasers with directional emission \cite{Wiersig.2006, Harayama.2015} or enhanced coupling \cite{Jiang.2017}. Phase-space analysis is a powerful tool to both describe and understand the optical modes in such deformed resonators, where the chaotic dynamics plays a major role. Instead of studying the real-space mode distribution, one looks at the field intensities as a function of both position and angle of incidence for a reduced subsystem, usually the resonator boundary. This method provides a thorough understanding of the underlying dynamics and is well suited to establish ray-wave correspondence \cite{Hentschel.2002}. A common phase-space representation are Husimi functions, which were introduced for open dielectric systems by Hentschel \textit{et. al.} \cite{Hentschel.2003} and have been used extensively to study the far-field patterns, wave dynamics and ray-wave correspondence. Recently, these phase-space approaches have been extended to include systems with non-homogenous refractive index \cite{Kim.2018} and have been used to study free space coupling into asymmetric cavities \cite{Shu.2013} and the dynamical evolution of light in a deformed cavity by calculating the respective functions at different times and following the evolution in both real space and phase space \cite{Chen.2019,Kwon.2013}. In this paper, we apply phase-space analysis based on generalized Husimi functions to study systems of coupled resonators, providing an intuitive understanding of the involved coupling processes. Coupled resonator systems have shown promising features for lasing systems \cite{Kreismann.2019} and non-hermitian physics \cite{Bosch.2019} and are as such a subject of interest. Firstly we present how the Husimi functions as derived in \cite{Hentschel.2003} can be used to study coupling phenomena in resonator systems and the difficulties involved. We then proceed to analyze the coupling in an illustrative waveguide-resonator system using the described method.
\section{Husimi functions for eigenmodes in dielectric cavities}
\begin{figure}
\subfigure[]{\raisebox{8mm}{
\includegraphics[width=0.17\textwidth]{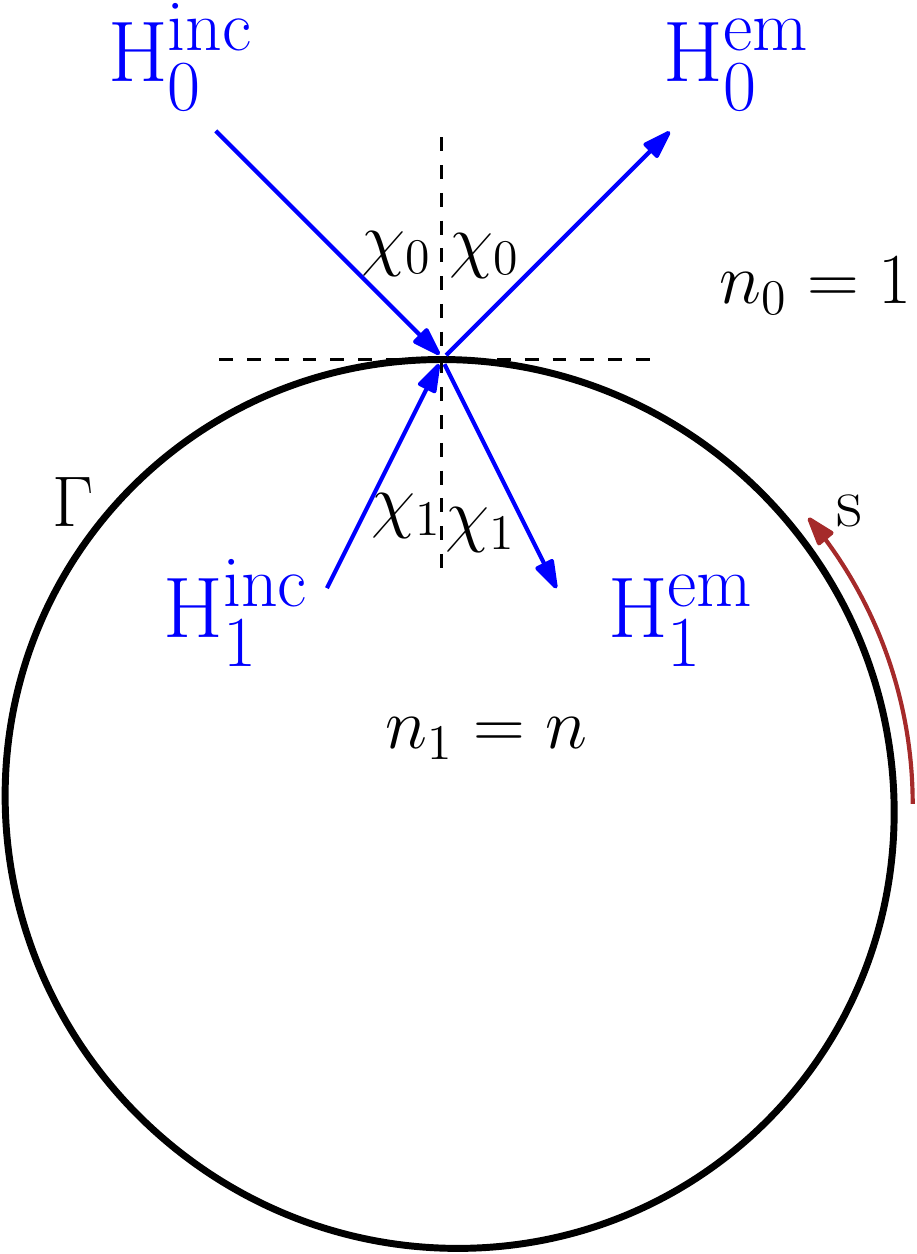}}
\label{fig:Skizze System}
}
\subfigure[]{\raisebox{8mm}{
\includegraphics[width=0.2\textwidth]{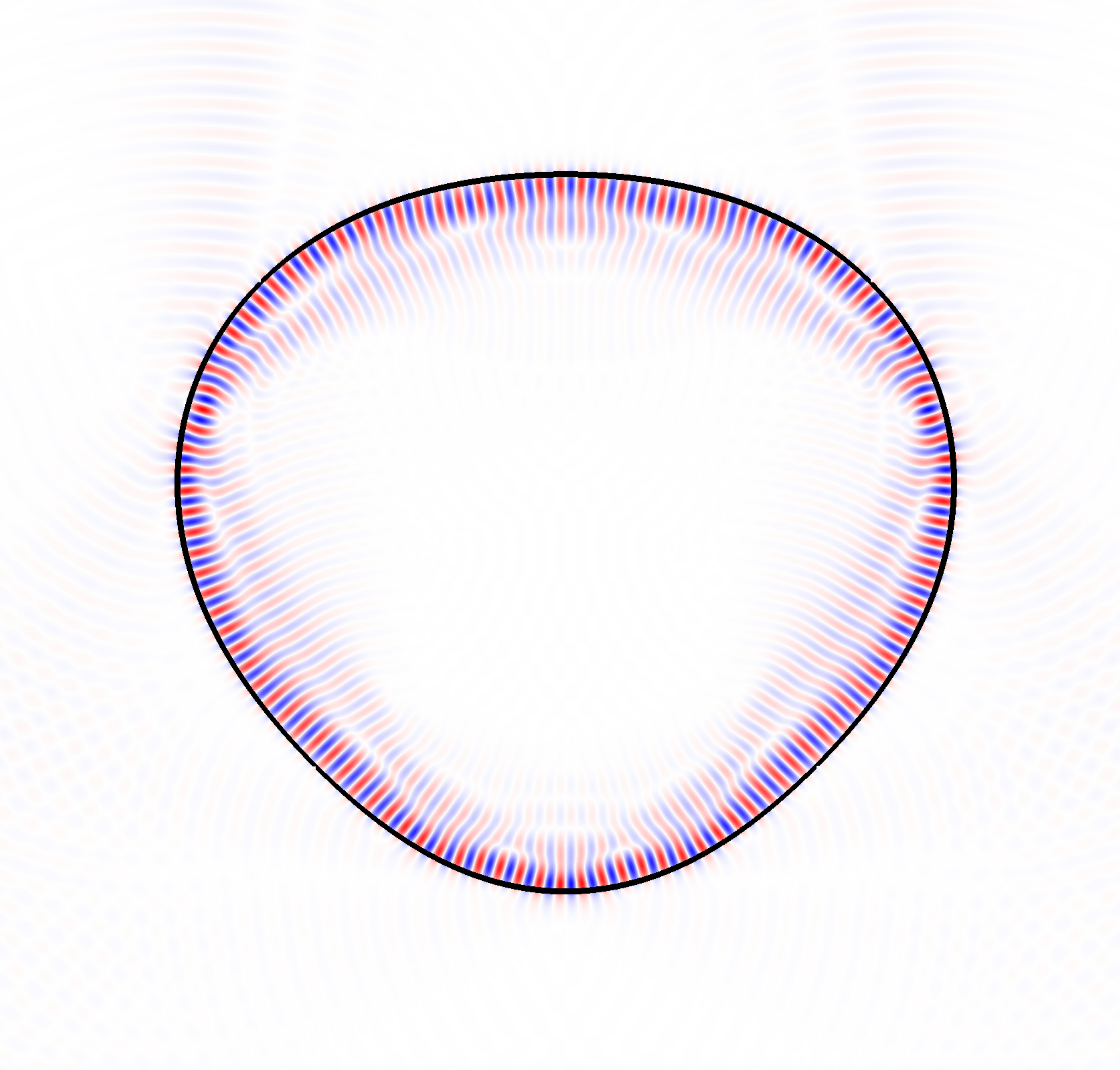}}
\label{fig:EF_Mode_1907}
}
\subfigure[]{
\includegraphics[width=0.6\textwidth]{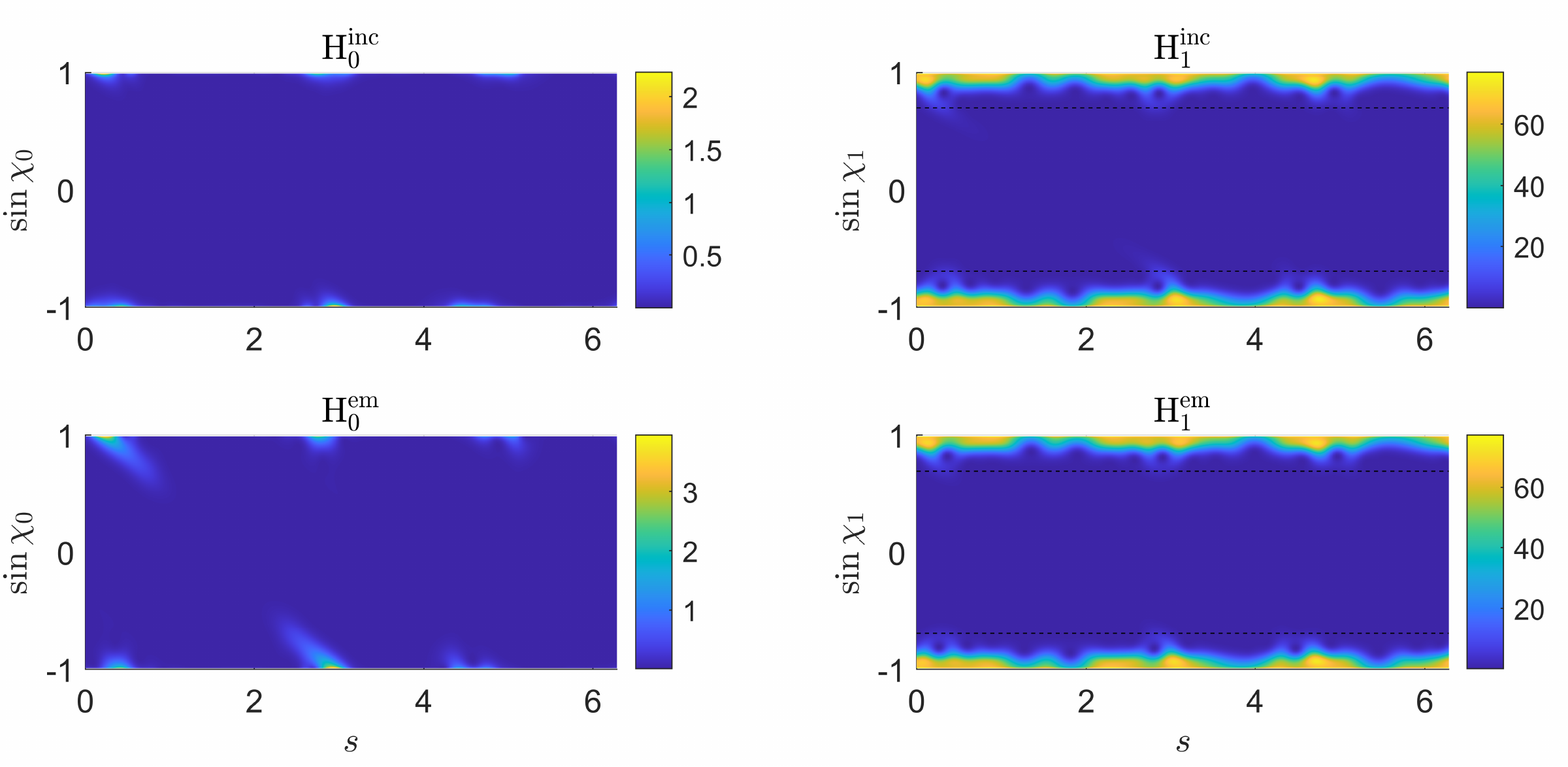}
\label{fig:Husimis_EF_Shortegg_Mode_1907}
}
\caption{(a) Incident and emerging rays at a dielectric boundary $\Gamma$ corresponding to the four Husimi functions and introduction of phase-space coordinates arc length $s$ (with $0 \leq s \leq 2 \pi)$ and angle $\chi$ of incidence. (b) Mode distribution of the shortegg cavity with $n = 1.44$ for a TE whispering-gallery eigenmode with the mode number $kR = 79.935 - 0.003i$. (c) The four Husimi functions calculated for the mode distribution shown in Fig.~\ref{fig:EF_Mode_1907}. Note that the scale depicts a relative intensity.}
\label{fig:Skizze_Mode_Husimi}
\end{figure}

The Husimi function was originally defined as a quasi-probability distribution in phase space \cite{Husimi.1940}, given by the overlap of the wavefunction with a Gaussian-type wavepacket (minimal-uncertainty). Applied to optical systems, it allows the representation of the field intensities as a function of position as well as momentum. Mathematically it is a windowed transformation, with the Gaussian window leading to the smallest possible uncertainty linked to such transformations. For 2D optical cavities, the Husimi function is usually calculated on the Poincaré surface of section (SOS) at the system boundary, leading to a phase-space representation of a reduced system which can be easily visualised in two dimensions \cite{Crespi.1993}. The values of the Husimi function correspond to the field intensities for a given boundary position $s$ and angle of incidence $\chi$ with respect to the boundary normal, where $\chi < (>) 0$ indicates (counter-) clockwise propagation direction. 
Hentschel \textit{et. al.} \cite{Hentschel.2003} presented four different Husimi functions for open dielectric systems with piecewise constant refractive indices, corresponding to the intensities of the incident (inc) and emerging (em) waves inside ($j=1$) and outside ($j=0$) of the interface (see Fig.~\ref{fig:Skizze System}). The four Husimi functions along the cavity boundary $\Gamma$ with Birkhoff coordinates $(s, \sin \chi)$ are defined as:
\begin{equation}
\label{eq:Husimi}
H_{j}^{inc(em)}(s, \sin\chi_j) = \frac{k_j}{2\pi} \left| {(-1)^j F_j h_j(s,\sin\chi_j) + (-) \frac{i}{k_0 F_j} h'_j(s,\sin\chi_j)}  \right|^2,
\end{equation}
with the weighting factors $F_j = \sqrt{n_j \cos{\chi_j}}$, the refractive indices $n_0=1$, $n_1$, the (vacuum) wave number $k_0$, and the overlap functions $h_j$, $h'_j$ given by
\begin{equation}
\label{eq:h}
h_j(s, \sin\chi_j) = \oint_\Gamma{ ds' \psi_j(s') \xi(s'; s, \sin\chi_j)},
\end{equation}
\begin{equation}
\label{eq:h'}
h'_j(s, \sin\chi_j) = \oint_\Gamma{ ds' \psi'_j(s') \xi(s'; s, \sin\chi_j)}.
\end{equation}
The wave functions $\psi$ (and its normal derivative $\psi'$) are taken on the respective side $j$ of the dielectric interface. The minimum-uncertainty wave packet $\xi$ is given by
\begin{equation}
\label{eq:chi}
\xi(s'; s, \sin\chi_j) = (\sigma \pi)^{-\frac{1}{4}} \sum_{l \mathsmaller{\in} \mathbb{Z}} \exp \left[ \frac{-(s' - s + 2\pi l)^2}{2\sigma} - ik_j \sin(\chi_j+2 \pi l) \right], 
\end{equation}
which is a periodic function in $s'$ centered around $(s', \sin\chi_j)$. The parameter $\sigma = \sqrt{2}/k_1$ determines the extension along the $s'$ direction and thereby the uncertainty in $\sin\chi_j$. The value $k_j$ is the wavenumber in each region, the angles of incidence are related by Snell's law $n_1 \sin\chi_1 = n_0 \sin\chi_0 $. This phase-space representation gives insight into the wave dynamics and allows the identification of regions in phase-space with high intensity. This approach has been used extensively to study asymmetric cavities \cite{Song.2010, Wiersig.2008, Shinohara.2010}. 

The four Husimi functions for an eigenmode of the shortegg cavity \cite{Schermer.2015} can be seen in Fig.~\ref{fig:Skizze_Mode_Husimi}. The whispering gallery mode (WGM) is confined by total internal reflection at the dielectric boundary. This can be seen in the Husimi functions inside the cavity $H^\mathrm{em}_1$ and $H^\mathrm{inc}_1$, which show high intensities for angles close to the boundary tangent only. The influence of the deformation can be seen in the deviation from perfectly straight lines, which is in turn the cause for the directional emission exhibited by this cavity. The differences between $H^\mathrm{inc}_1$ and $H^\mathrm{em}_1$ can be seen clearly for the points in the leaky region (between the dashed lines $\sin \chi_{cr} = \pm 1/n $), for which the light leaves the cavity according to Snell's law. Outside these lines, total reflection occurs and the incoming and emitted components are similar. The high intensity emission points along the boundary can be identified in $H^\mathrm{em}_0$ and $H^\mathrm{inc}_1$. The directional emission can be seen clearly as well in phase-space: the highest outgoing intensities outside of the cavity in $H^\mathrm{em}_0$ are localized along a line joining the positions $(s = 0.35, \sin\chi_0 = 1)$ and $(s = \pi - 0.35, \sin\chi_0 = -1)$. This closely resembles the signature of a plane wave emerging from $s \approx \pi/2$, in accordance with the shortegg's farfield emission characteristics.\\ 
%There are also some lower intensity regions corresponding to a backwards emission from the same points and a two-sided emission localized around the bottom of the cavity ($s=3\pi/2$).\\
%
\section{Husimi functions for waveguide-resonator coupled systems}
\begin{figure}
\subfigure[]{\raisebox{5mm}{
\includegraphics[width=0.26\textwidth]{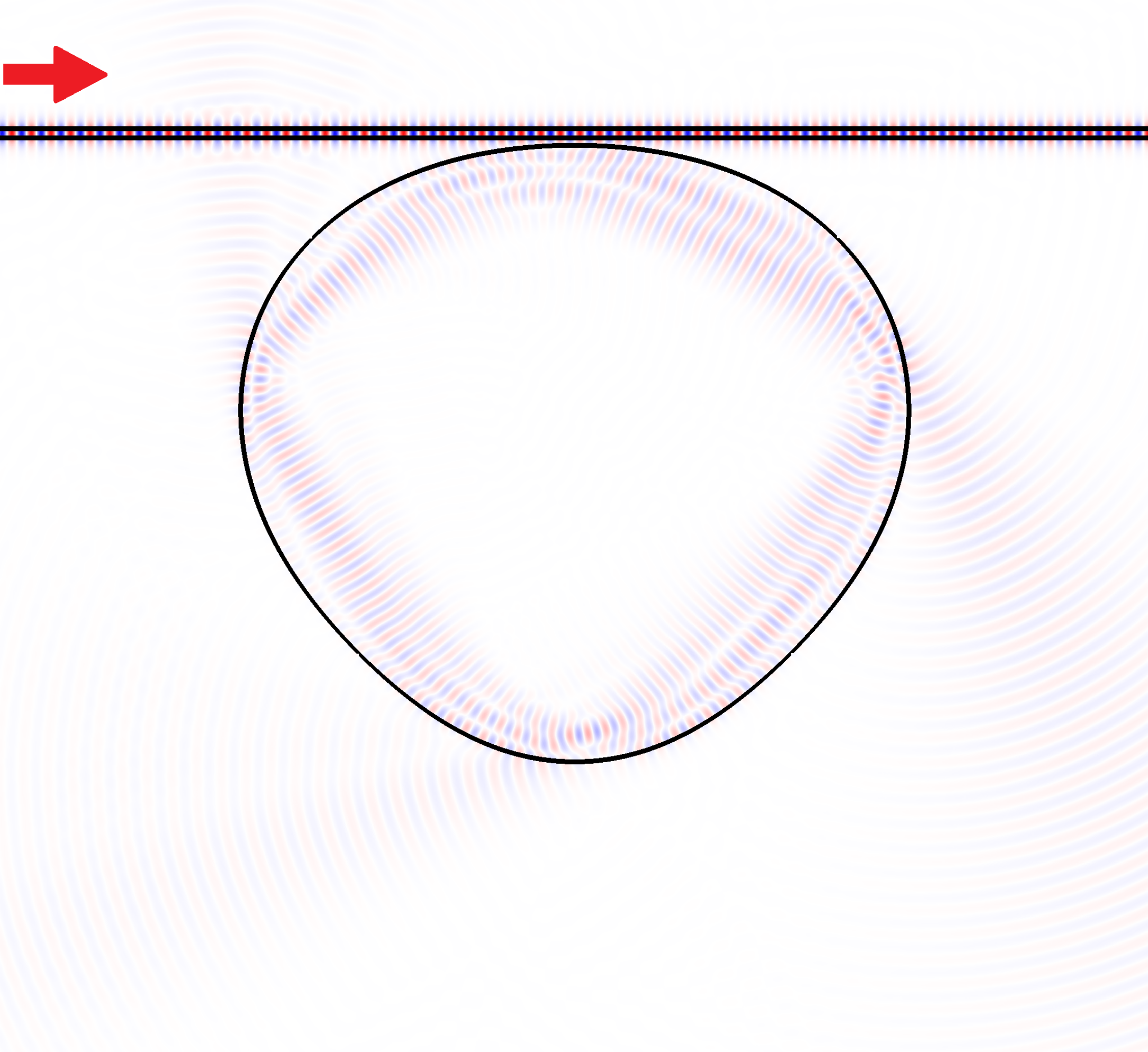}}
\label{fig:Full_Coupled_Mode_off}
}
\subfigure[]{
\includegraphics[width=0.7\textwidth]{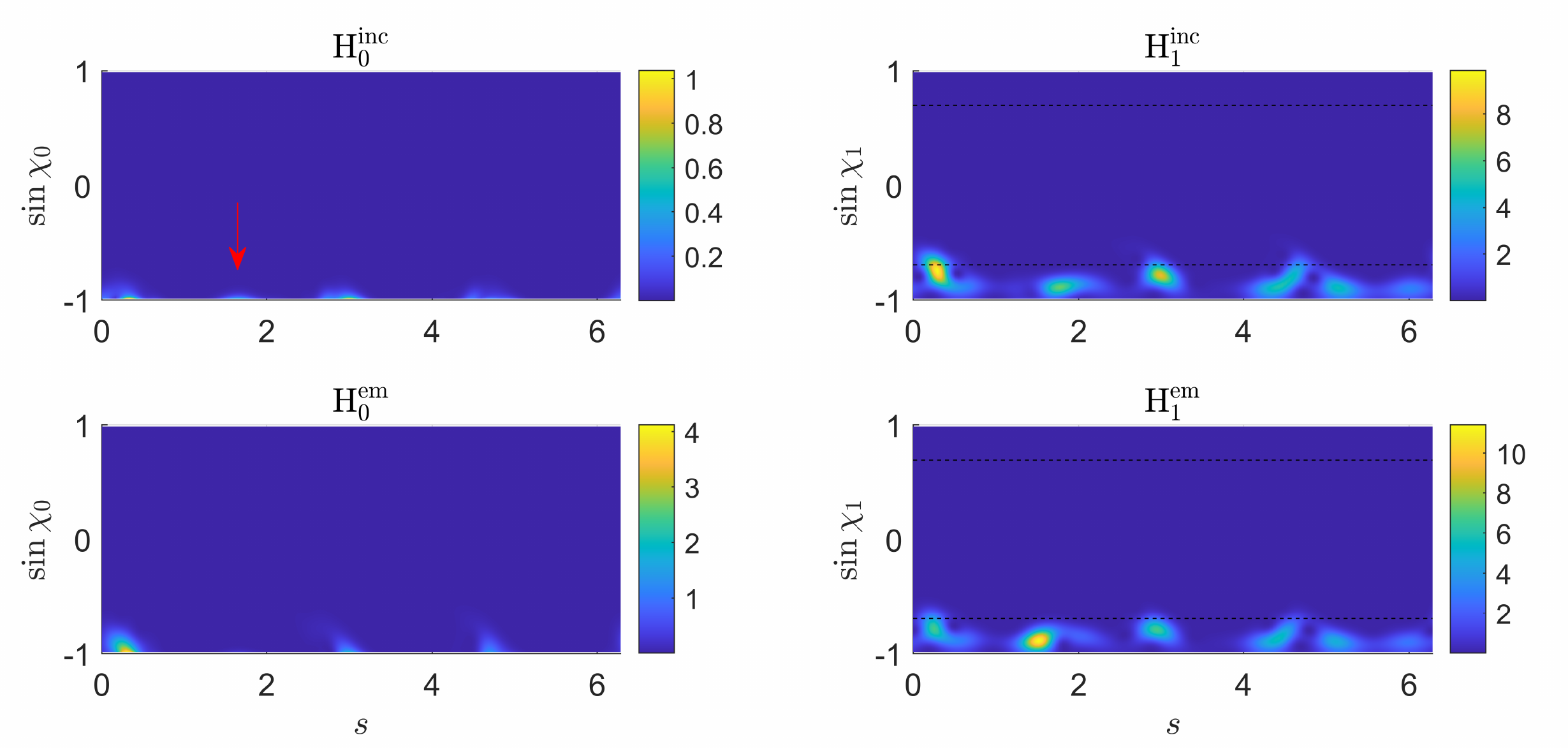}
\label{fig:Full_Coupled_Husimi_off}
}\\
%-----------------------------------
\subfigure[]{\raisebox{5mm}{
\includegraphics[width=0.26\textwidth]{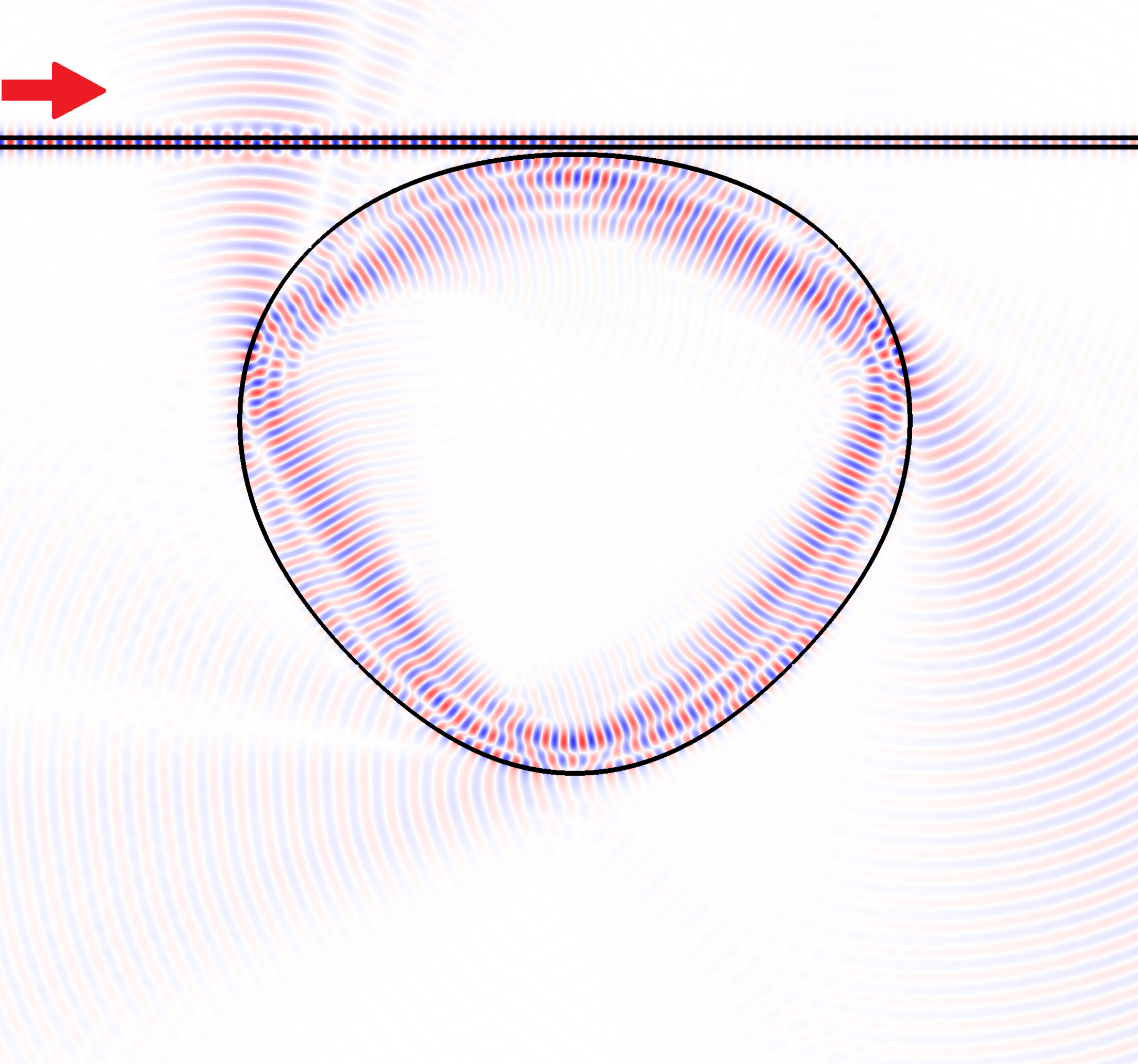}}
\label{fig:Full_Coupled_Mode_res}
}
\subfigure[]{
\includegraphics[width=0.7\textwidth]{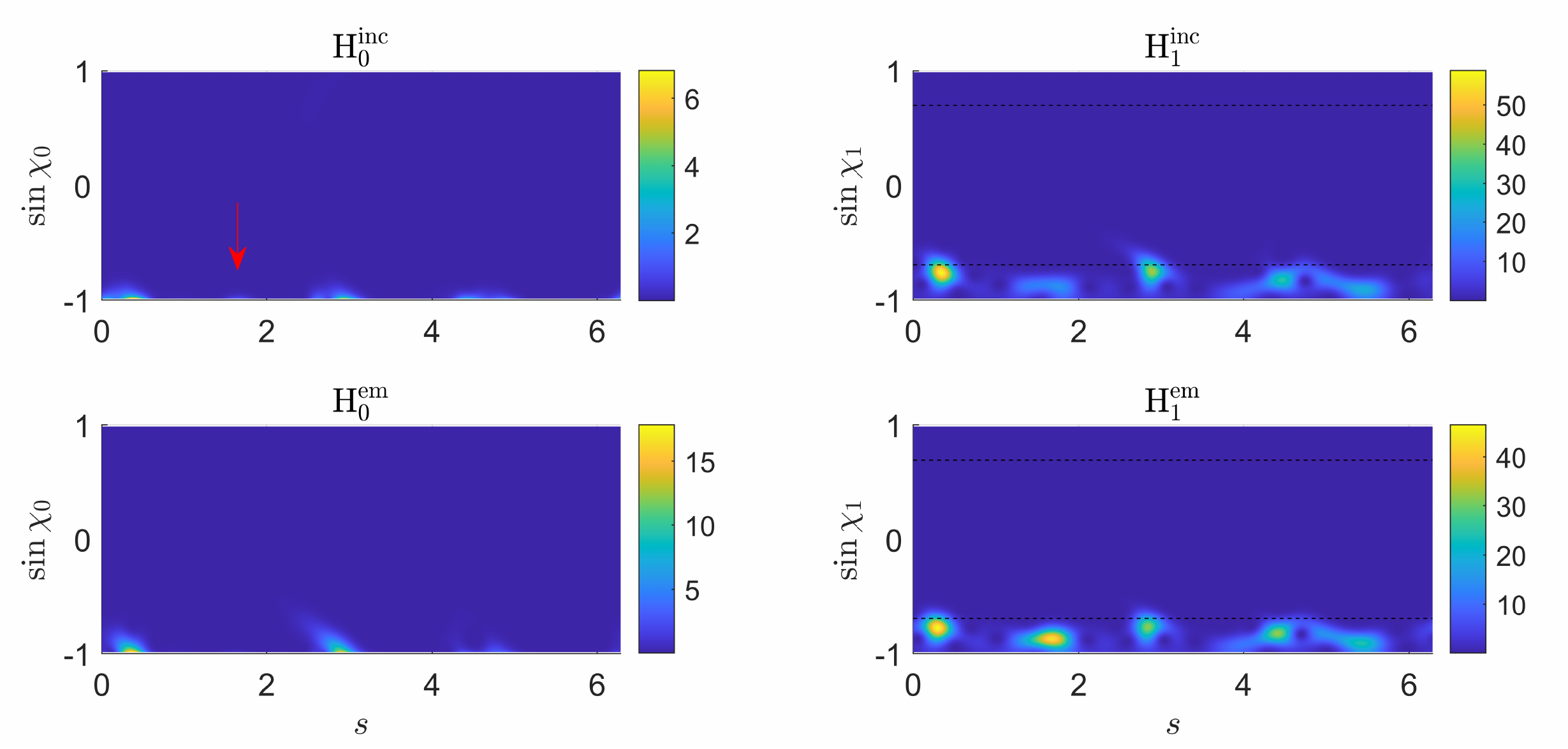}
\label{fig:Full_Coupled_Husimi_res}
}
%-----------------------------------
\caption{ (a) Mode distribution of the waveguide-shortegg system at the off-resonant frequency $\mathrm{Re}(kR) = 80.015$. (b) Husimi functions corresponding to the mode showed in (a). (c) Mode distribution of the waveguide-shortegg system at the resonant frequency $\mathrm{Re}(kR) = 80.211$. (d) Husimi functions corresponding to the mode showed in (c). The width of the waveguide is set to $w_{WG} = 0.565 \lambda / n_{in}$ and the distance between the resonator and the waveguide is $\Delta = \lambda /2$. The waveguide has the same refractive index as the cavity $n_{in}=n=1.44$.}%
\label{fig:fullCoupled}
\end{figure}

When analyzing coupling processes, it is necessary to look at whether the incoming fields excite the resonant mode at the right position as well as with the right angle of incidence. The phase-space analysis of coupling problems via Husimi functions allows to do this naturally by identifying the signatures of the $H^\mathrm{inc}_0$ function for the resonator of interest in a coupled system. $H^\mathrm{inc}_0$ gives insight into how incoming light from neighbouring regions behaves at the resonator boundary and can thus provide an explanation for the coupled system response. The excitation of supported modes is expected to be directly correlated to the overlap between the incoming exciting fields and the resonant eigenmode. A problem with this approach, however, arises due to the relative intensities of the incoming components and the resonant fields. Fig.~\ref{fig:fullCoupled} shows the Husimi functions for a coupled waveguide-shortegg system at an off-resonant as well as for a resonant frequency. The fields in the resonator were excited by an incoming field distribution from the left (red arrow) calculated from a port boundary analysis. The incoming field components $H^\mathrm{inc}_0$ from the waveguide are expected to be located around ($s=\pi/2, \, \sin \chi_0 = -1$, see red arrow), but they can barely be distinguished from the intensities arising due to the intra-cavity field in the resonant case. In the off-resonant case, the waveguide component can be seen in $H^\mathrm{inc}_0$ due to the lower intra-cavity intensity.

For all practical purposes, representing the fully coupled waveguide-resonator system in phase space will cause difficulties related to the dominance of the resonant mode in all the Husimi functions. 
% The existence of this unexpected $H^{inc}_0$ components is tied to the used windowed transformation, which causes an uncertainty in $\sin\chi_j$. This complicates the identification of incoming field components in phase-space, especially when looking at resonant modes for which the intra-cavity intensity is greatly enhanced. 
One could try to change the Gaussian pulse in order to attain a higher precision in $\sin\chi_j$, but this is only possible through an undesired trade-off with increasing the uncertainty in $s$.\\
%, hereby making the trade-off useless for practical purposes.\\
%
To overcome this issue, one can compute the field components originating from the coupled system (waveguide, neighbouring resonator) separately.
%and perform a phase-space breakdown WAS IST DENN DAS??? 
In particular, one can compute the waveguide's field distribution at the resonator boundary without including the resonator in the calculation, hereby preventing the resonant intensity from overshadowing the analysis of the coupling. To this end, the Husimi projection is taken at the boundary where the resonator to be coupled into would be situated. This result is subsequently used to calculate the overlap with the resonant mode's Husimi function (using the Husimi function $H_1^\mathrm{em}$ in both cases). 

Although Husimi functions were originally motivated and derived for dielectric boundaries, it is not restricted to this, and the overlap of the wave field with a minimum-uncertainty wave packet can be calculated along an arbitrary boundary $\Gamma$. 
%but by looking at equation \eqref{eq:Husimi} one can see that it is mathematically a Gabor transformation of the field along a certain boundary $\Gamma$. 
This allows us to proceed in calculating the incoming and outgoing field components originating at the waveguide mode along the boundary where the resonator would be while setting $n_{0} = n_{1} = 1$. The overlap $S_{\mathrm{Exc,Res}}$ between the Husimi functions of the exciting field $H^\mathrm{em}_{1,\mathrm{exc}}$ and the Husimi Functions of a chosen resonant mode $H^\mathrm{em}_{1,\mathrm{res}}$ can be calculated via 
\begin{equation}
\label{eq:OverlapStrength}
S_{\mathrm{Exc,Res}} = \frac{\int \int ds \, \, d\sin \chi_j H^{em}_{1,\mathrm{exc}}(s,\sin \chi_j) H^{em}_{1,\mathrm{res}}(s,\sin \chi_j) }{\int \int ds \, \, d\sin \chi_j}
\end{equation}
and is correlated to the coupling strength between the incoming field and the resonant mode. Note that both $H^{em}_{1,\mathrm{exc}}$ and $H^\mathrm{em}_{1,\mathrm{res}}$ have to be evaluated along the same boundary $\Gamma$.

The results are not expected to match full-wave calculations perfectly due to two reasons: first, the uncertainty arising from the Husimi projection and second, the fact that interactions between the cavity and the exciting system can not be accounted for. In the weak coupling regime, realized if the distance between the resonator and the waveguide is large enough, the eigenmodes of the coupled resonator remain essentially unchanged. Consequently, the results should correlate well with full numerical wave simulation results of the combined system. More importantly, this analysis gives an intuitive understanding of the involved coupling processes in phase space. If one is interested in a time-dependent analysis, the phase-space description of the incoming fields can be used as time-dependent input for the computation of the dynamical fields inside the resonator.\\
\begin{figure}
\subfigure[]{\raisebox{5mm}{
%\centering
\includegraphics[width=0.28\textwidth]{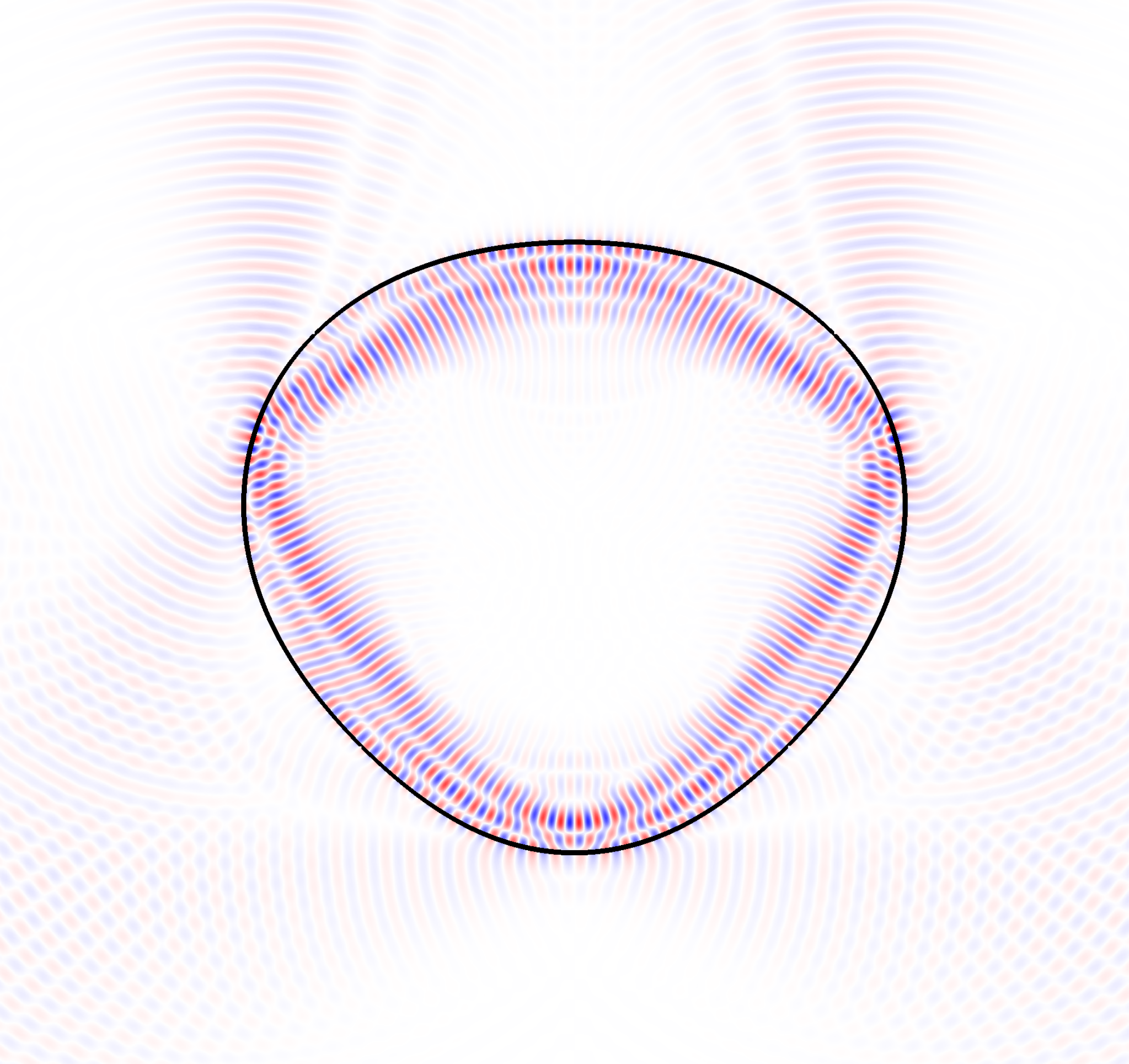}}
\label{fig:H_ModeDistribution}
}
\subfigure[]{
%\centering
\includegraphics[width=0.7\textwidth]{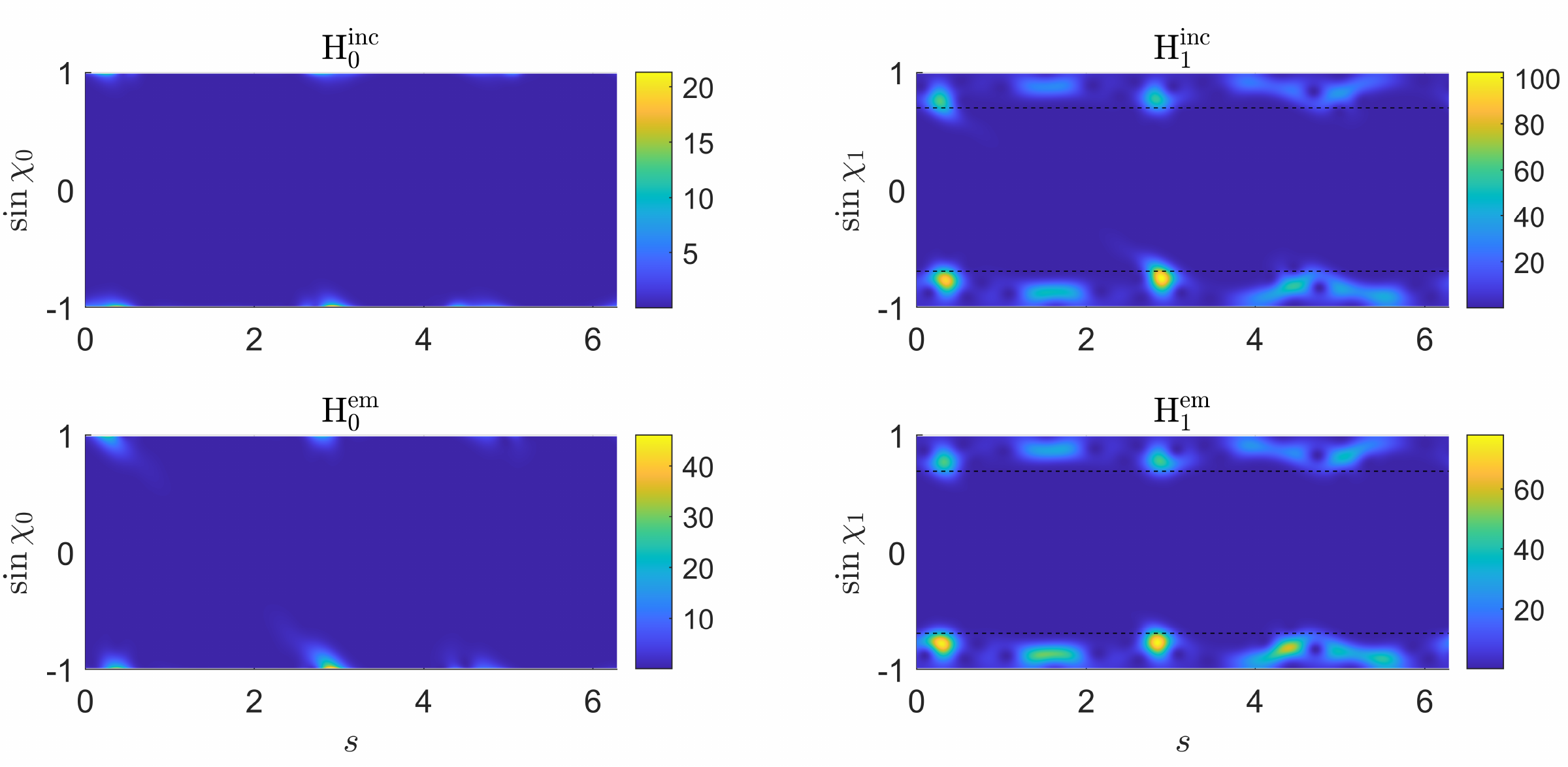}
\label{fig:Husimis_EF_Shortegg}
}
\caption{(a) Mode distribution in the shortegg cavity with $n = 1.44$ for a TE eigenmode with the mode number $kR = 80.004 - 0.029i$.  (b) The four Husimi functions calculated for the mode distribution showed in \ref{fig:H_ModeDistribution}, with the dashed line marking the critcal angles $\sin \chi_i = 1/n$. The Husimi functions inside show four areas with high intensity along $s$, which can be linked to the four reflection areas on which the mode distribution along the boundary has the highest values. %The difference between $H^{inc}_1$ and $H^{em}_1$ can be seen mainly above the critcal line due to transmission effects. $H^{em}_0$ correlates well with the far-field emission positions seen in the mode distribution.
}
\label{fig:EF_Husimi}
\end{figure}
In order to illustrate this method, we studied the transmission through the waveguide in dependence on the orientation angle of the shortegg resonator and compared the results 
%angle dependent transmission spectra for the asymmetric shortegg cavity using
from conventional numerical methods for the full system with the phase-space analysis described above. The electric field distributions used for the eigenmodes, the phase-space analysis as well as the transmission curves were calculated by solving the 2D Helmholtz equation in the frequency domain, with outgoing wave conditions imposed at infinity \cite{Jackson.1998}.
%\begin{equation}
%\label{eq:Helmholtz}
%- \nabla^2 \psi(x,y) = n^2(x,y) k^2 \psi(x,y) 
%\end{equation}
%where $k=\omega/c$. The harmonic solutions for the electric field are given by $E_z(x,y) = \psi(x,y) e^{-i\omega t}$. 

We focus here on TE modes, where $E_\mathrm{z}$ and its normal derivative are continuous across dielectric boundaries. The fields were solved for numerically in COMSOL v5.4 \cite{Comsol.2018}, using perfectly matched layers (PMLs) at the edges of the truncated system. %For the transmission spectra we used numerical ports. 
The analyzed system consisted of a waveguide coupled to a shortegg cavity, whose geometric shape is given, in polar coordinates $(r, \phi)$, as 
%radius is given by 
%\begin{equation}
%\label{eq:RadiusShortegg}
$r(\phi) = R_0(1 + 0.16 \cos(\phi) - 0.022 \cos(2\phi) - 0.05 \cos(3\phi)).$
%\end{equation}
For both the waveguide and the shortegg cavity the refractive index was set to $n=1.44$. The distance between the resonator and waveguide was held constant at $\Delta = 0.5 \lambda$ for all angles by shifting the center of the resonator. 

As known from in-house experimental data \cite{Behrens.2020}, the transmission spectrum (i.e., intensity transmitted through the waveguide as a function of the orientation angle) is highly dependent on the orientation of the waveguide relative to the resonator. Although this is to be expected due to the asymmetry of the resonator, an accurate, quantitative, and predictive explanation is accessible by analyzing the system in phase space. 
%For this we first looked at the resonant mode and the angle dependent transmission was calculated. 

\begin{figure}
\begin{center}
\subfigure[]{\raisebox{5mm}{
%\centering
\includegraphics[width=0.3\textwidth]{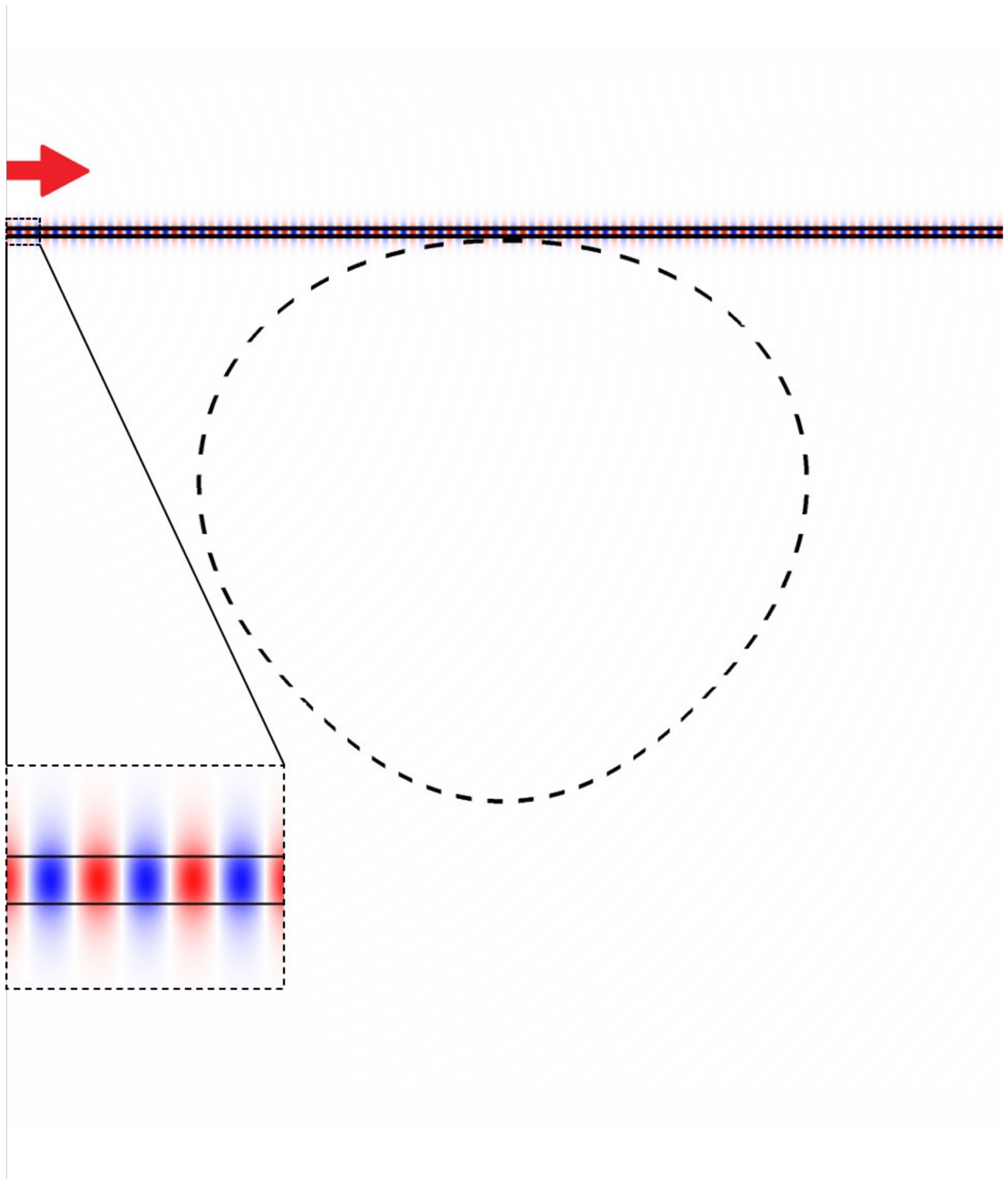}
\label{fig:dashed_mode_wg}
}}
\subfigure[]{
%\centering
\includegraphics[width=0.48\textwidth]{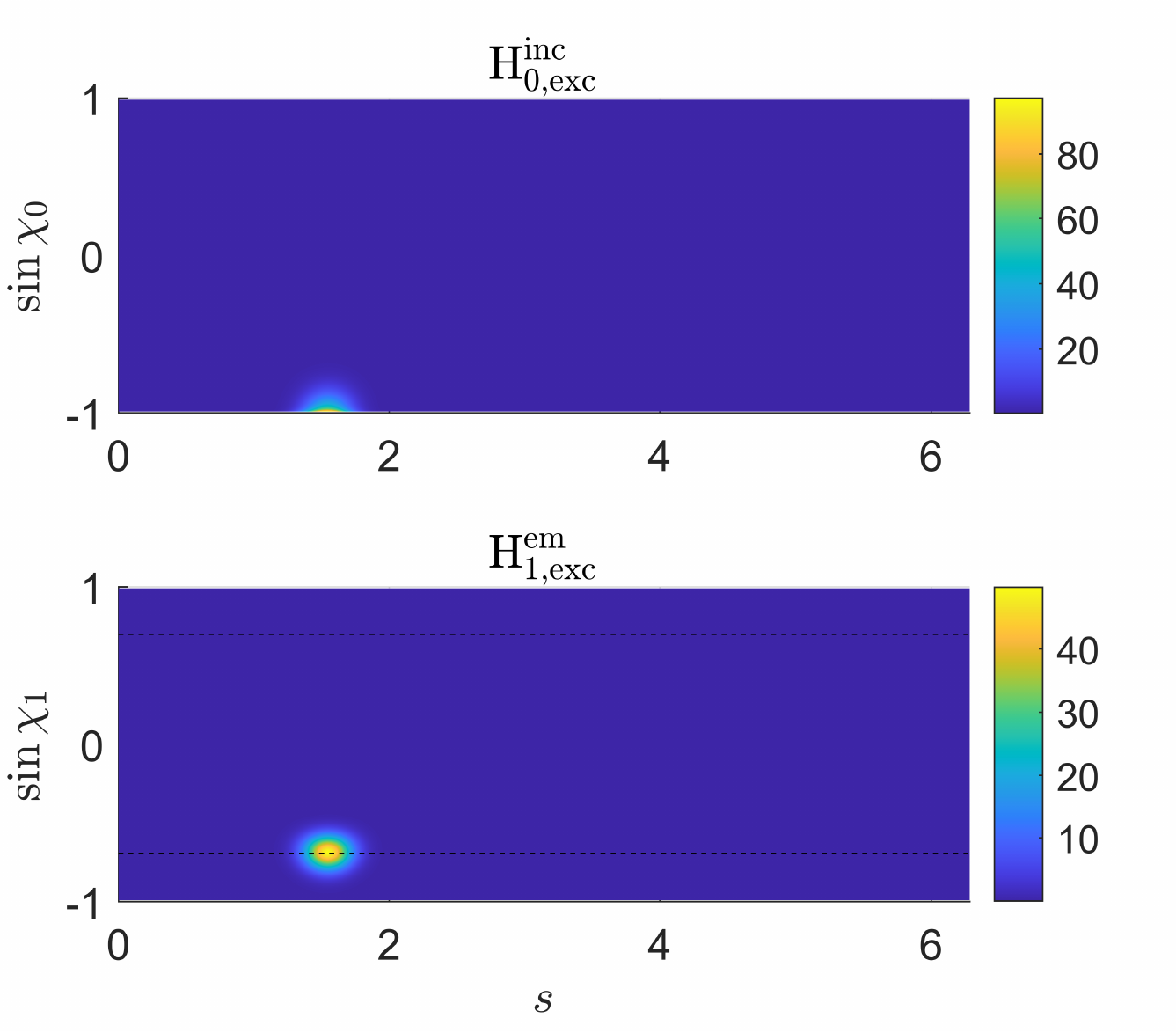}
\label{fig:Husimis_Dashed_Snell}
}
\\
\subfigure[]{{
%\centering
%\includegraphics[width=0.45\textwidth]{Transmission_Comp.png}
%\input{Transmission360deg_2Axis.tex}}
\includegraphics[width=0.4\textwidth]{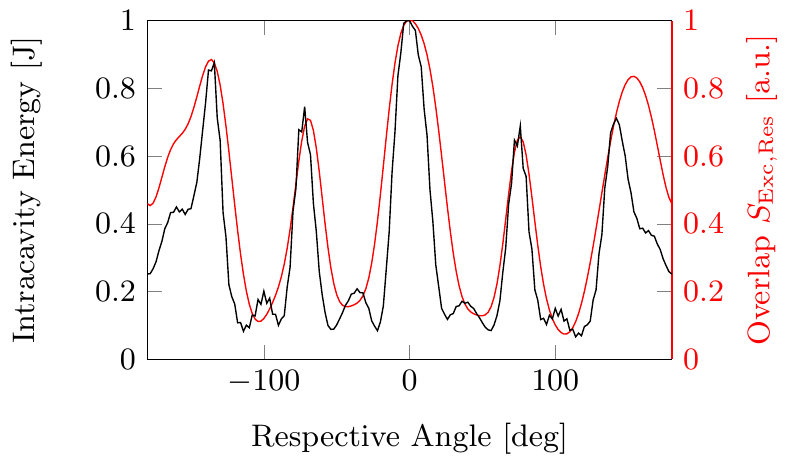}
}
\label{fig:Intensity_Comparision}
}
\subfigure[]{{
%\centering
%\includegraphics[width=0.45\textwidth]{Transmission_Comp.png}
%\input{Transmission360deg_2Axis.tex}}
\includegraphics[width=0.4\textwidth]{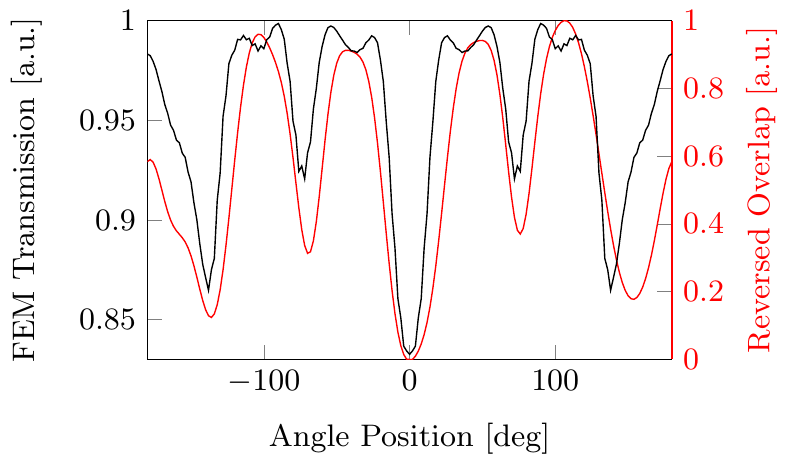}
}
\label{fig:Transmission_Comparision}
}
\caption{(a) Waveguide-only mode used to compute the exciting fields on the boundary marked by the dashed lines. (b, top) Husimi function $H^\mathrm{inc}_{0,\mathrm{exc}}$ computed from the field distribution in (a) on the dashed boundary. (b, bottom) $H^\mathrm{em}_{1,\mathrm{exc}}$ inside the resonator boundary computed by applying Snells and Fresnel laws to $H^\mathrm{inc}_{0,\mathrm{exc}}$ on (b,top). (c) Comparison of the normalized intracavity energy and the overlap $S_{\mathrm{Exc,Res}}$ for the waveguide-shortegg coupled system. (d) Comparison of the transmission and the reversed normalized overlap for the same system. The black curves (left axis) are obtained from the full-wave simulation, the red curves (right axis) from the phase space analysis.}
\end{center}
\end{figure}

A typical mode distribution as well as the four Husimi functions can be seen in Fig. \ref{fig:H_ModeDistribution}. Note that the resonant frequency for the coupled system shown in \ref{fig:fullCoupled} and the eigenfrequency vary slightly due to the coupling.
The values $H^\mathrm{em}_{1,\mathrm{res}}$ are computed from eigenvalue calculations. To calculate the fields used for $H^\mathrm{em}_{1,\mathrm{exc}}$ originating at the waveguide, the mode distribution of the waveguide-only system (see Fig. \ref{fig:dashed_mode_wg}) is computed. The waveguide mode corresponds to the fundamental mode for the used frequency and waveguide width and was derived from a numerical boundary mode analysis.  
%HIER MEHR DETAILS ZUR WAVEGUIDE-MODE!
%. This values correspond to

Figure \ref{fig:Husimis_Dashed_Snell} shows its Husimi function $H^\mathrm{inc}_0$ at the boundary marked by the dashed line (top), from which the refracted field can be computed using Snell's and Fresnel's laws (bottom) yielding $H^\mathrm{em}_{1,\mathrm{exc}}$ used in Eq. \eqref{eq:OverlapStrength}. The overlap value $S_{\mathrm{Exc,Res}}$ can be computed for each orientation angle by shifting the center of $H^\mathrm{em}_{1,\mathrm{exc}}$ to the respective angle. This method requires considerably less computational effort than computing the transmission for each orientation angle individually via a full-wave simulation, since here the eigenfrequency and the excitation field have to be computed only once from FEM simulations instead of for each angle separately. The  dependence on the orientation angle is accounted for in the overlap computation. Figure.~\ref{fig:Intensity_Comparision} shows the intracavity energy computed with FEM as well as the overlap $S_{\mathrm{Exc,Res}}$ calculated from the phase-space analysis as function of the respective angle, which is $0^o$ for the resonator position depicted in Fig.~\ref{fig:dashed_mode_wg}. In experiments it is common to measure the transmission, since the intracavity energy is not easily accessible and both values are connected due to energy conservation. Note that a big overlap value corresponds to a strong excitation of the mode and therefore a low transmission through the waveguide, thus the overlap values are normalized and substracted from unity (reversed overlap) to yield a comparative measure for the waveguide transmission: $S_{rev}(\theta) = 1 - S_{\mathrm{Exc,Res}(\theta)}/\max(S_{\mathrm{Exc,Res}})$, as displayed in Figure.~\ref{fig:Transmission_Comparision}.

In Fig.~\ref{fig:Transmission_Comparision} one can observe the missing symmetry about the $0^o$-position of the reversed overlap value and the FEM-computed transmission. This is linked to observing only in-coupling components into the cavity and not correcting by how the cavity fields couple back into the waveguide. A close look at $H^\mathrm{em}_{1}$ in Fig.~\ref{fig:Husimis_EF_Shortegg}, shows that the Husimi function is not perfectly symmetric around $(s=\pi/2, \, \sin \chi_1 = -0.8)$, leading to the asymmetry displayed here. As expected, this effects are not relevant when looking at the intra-cavity energy. Nonetheless, the results agree semi quantitatively with the values from full-wave simulations. The differences can be attributed to the uncertainty involved in the calculation of the Husimi function as well as not having corrected for the non-constant curvature of the shortegg cavity. This could be accounted for by computing $H^\mathrm{em}_{1,\mathrm{exc}}$ for each angle, but it comes with a considerable computational effort and thus negates one of the advantages of the used method, since the used method the computation involves only the evaluation of an overlap integral and it provides a reasonable approach to full numerical simulations.

%In this work, we show how a phase-space approach can be used to analyze systems of coupled dielectric resonators. While a direct analysis of the coupling behaviour via (incoming) Husimi functions is not feasible in the case of strongly coupled systems, an alternative method valid for the weak coupling regime was introduced.

%
\section{Summary}

In this work we introduce a phase-space approach based on Husimi functions to analyze systems of coupled dielectric resonators. The method involves the computation of the incoming field components at the boundary where the coupling occurs for the individual non-coupled subsystems. From this we compute the respective Husimi functions and use their overlap to characterize the coupling efficiency. The method was verified on a system consisting of an asymmetric resonator excited by coupling to a waveguide. The proposed method provides a good approximation of full FEM calculations with considerably lower computational effort. In addition, this analysis gives intuitive insights into the coupling mechanisms taking place at the cavity boundaries. This method can be applied to more complex coupled optical systems consisting of multiple optical components, enabling a deeper understanding of the coupling processes and an effective approximation of the expected behaviour for the weak coupling regime.

%%%%%%%%%% If using BibTeX:
\bibliography{PhaseSpace_Bibliography}

\end{document}